# Using Optimal Ratio Mask as Training Target for Supervised Speech Separation

Shasha Xia, Hao Li and Xueliang Zhang
Inner Mongolia University, Hohhot, China
E-mail: cszxl@imu.edu.cn  Tel: +86-0471-4993132

*Abstract*—Supervised speech separation uses supervised learning algorithms to learn a mapping from an input noisy signal to an output target. With the fast development of deep learning, supervised separation has become the most important direction in speech separation area in recent years. For the supervised algorithm, training target has a significant impact on the performance. Ideal ratio mask is a commonly used training target, which can improve the speech intelligibility and quality of the separated speech. However, it does not take into account the correlation between noise and clean speech. In this paper, we use the optimal ratio mask as the training target of the deep neural network (DNN) for speech separation. The experiments are carried out under various noise environments and signal to noise ratio (SNR) conditions. The results show that the optimal ratio mask outperforms other training targets in general.

## I. INTRODUCTION

Speech separation aims to extract the target speech signal from a noisy mixture and it is meaningful for many applications such as robust automatic speech recognition (ASR) and hearing aids design. Monaural speech separation is the most common case and also very challenging, because only one channel signal could be used. In this study, we focus on monaural speech separation. This problem has been studied for many years. The early methods, e.g. spectral subtraction [1], have stationary assumptions on background noise, which limit their application. Computational auditory scene analysis (CASA) [2] stimulates human auditory mechanism and employs an Ideal binary mask (IBM) [3] as the computational goal. The IBM indicates the target speech on a time-frequency (T-F) representation, where "1" and "0" indicates a T-F unit dominated by target speech and noise respectively. With this concept, it is natural to formulate the speech separation as a binary classification problem which can be solved by supervised learning algorithms.

For the supervised speech separation, three key factors are mainly concerned, i.e. learning machines, features and training targets. Several learning machines have been investigated in the literature, e.g. Gaussian mixture model (GMM) [4], support vector machine (SVM) [5] and Deep Neural Networks (DNN) [6]. Typically, DNN-based speech separation has achieved a great success, because its strong learning capacity enables effective modeling of nonlinear interactions between speech and the acoustic environments as well as dynamic structure of speech. Discriminative features are also important. In [7-8], extensive features are studied in noisy and reverberant conditions.

For the training target, we should keep in mind that the ideal target can obtain good separation results, and its estimation shouldn't be hard by current learning machines. The IBM and Ideal Ratio Mask (IRM) [9] are commonly used targets. Separating with the IBM usually introduces residual musical noise and speech quality cannot be improved. IRM can be thought of as smooth form of IBM, which improves both speech quality and intelligibility. Recent works show that phase information is also important for speech separation. Based on these researches, complex Ideal Ratio Mask (cIRM) [10] and Phase Sensitive Mask (PSM) [11] were proposed. The cIRM is computed on complex domain. Although the cIRM can perfectly recover the target speech, it increases the difficulty of estimation. The PSM introduces the phase information and operates on real domain.

Recently, Optimal Ratio Mask (ORM) is proposed in [12], which can be viewed as an improved version of the IRM. The ORM considers the correlation between the target speech and noise in real environment. Theoretical analysis shows that the ORM can improve the SNR over the IRM. However, is the ORM a good training target for supervised speech separation algorithm? This key question, as we mentioned earlier, is not answered.

This paper aims to investigate the performance of ORM when applied in DNN-based monaural speech separation. The rest of this paper is organized as follows: next Section describes the framework and procedure of DNN-based monaural speech separation system. In Section III, we describe the calculations of five training targets adopted for comparison. The results are shown in Section IV. We conclude the paper in the last Section.

## II. DNN-BASED MONAURAL SPEECH SEPARATION

The framework of the DNN-based monaural speech separation used in this study is same as in [9]. We use a set of complementary features consisting of amplitude modulation spectrogram (AMS), relative spectral transform and perceptual linear prediction (RASTA-PLP) and Mel-frequency cepstral coefficients (MFCC). The feature set used here is similar to the one in [9]. Since useful information is carried across time frames, a symmetric 5-frame context window is used to splice adjacent frames into a single feature vector.

The DNN is composed of three hidden layers, each layer has 1024 rectified linear hidden units (ReLU) [13]. The back propagation with dropout regularization (dropout rate 0.2) [14] is used for network training. Adaptive gradient descent [15] coupled with a momentum term as the optimization technique. Momentum rate is 0.5 during first 5 epochs and 0.9 during the rest epochs. The goal of network training is to output an ideal estimation of training target. We use the mean squared error (MSE) as cost function. The number of output units is correspond to the dimensionality of the training target. The sigmoid activation function is applied when target is in the range of [0, 1], otherwise linear activation function is applied. The general architecture is shown in Fig. 1.

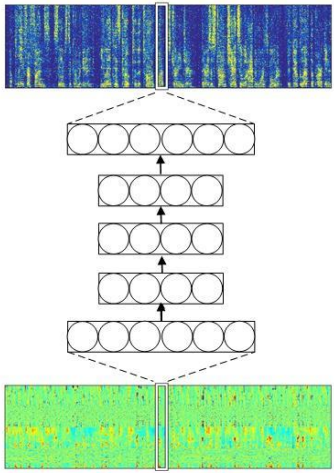

Fig. 1. The architecture of DNN-based speech separation.

### III. THE OPTIMAL RATIO MASK

The definition for the ORM is derived by minimizing the mean square error (MSE) of clean speech and estimated target speech [12].

$$\gamma(t,f) = \frac{|S(t,f)|^2 + \mathcal{R}(S(t,f)N^*(t,f))}{|S(t,f)|^2 + |N(t,f)|^2 + 2\mathcal{R}(S(t,f)N^*(t,f))} \quad (1)$$

where, $S(t,f)$ and $N(t,f)$ are the spectrum of speech and noise at frame $t$ and frequency $f$. $\mathcal{R}(\circ)$ denotes the real component of spectrum and '*' denotes the conjugate operation.

It can be seen that ORM is very similar to IRM, except for the coherent part $\mathcal{R}(S(t,f)N^*(t,f))$, which is assumed to be 0 in IRM. In fact, this assumption is too strong. Figure 2 shows the spectral coherence between the speech and the noise in a noisy speech. We can see that the speech and the noise are highly correlated. The ORM is proven to get better performance for speech separation in [12].

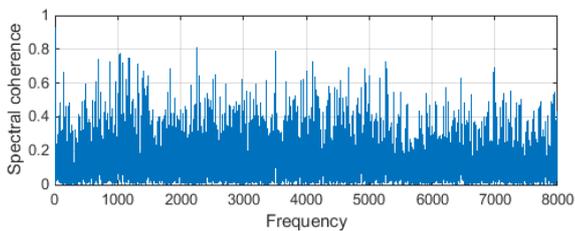

Fig. 2. Coherence estimation of the speech and noise signals in a noisy speech.

The ORM varies in the range of $(-\infty, +\infty)$ that is not easy to estimate. So, we restrict the value of the ORM with a hyperbolic tangent as in [9]

$$\text{ORM}(t,f) = K \frac{1 - e^{-c\gamma(t,f)}}{1 + e^{-c\gamma(t,f)}} \quad (2)$$

where, $c = 0.1$ is steepness and $K=10$ restricts the range of the ORM to $(-10, +10)$. $\gamma(t,f)$ is the original ORM defined in eq. (1).

### IV. VARIOUS TRAINING TARGETS

In this section, we will introduce several mask targets. The separation system employing mask target is called mask-based speech separation.

#### A. Ideal binary mask (IBM)

The IBM the simplest mask which is defined as following.

$$\text{IBM}(t,f) = \begin{cases} 1, & \text{if } |S(t,f)|^2 - |N(t,f)|^2 > \theta \\ 0, & \text{otherwise} \end{cases} \quad (3)$$

where $\theta$ is the local threshold in T-F units, and it is set to zero.

#### B. Ideal ratio mask (IRM)

From eq. (3), we can see that the IBM makes a hard decision according to the energy of speech and noise in T-F unit. IRM can be viewed as a soft decision, which is defined as,

$$\text{IRM}(t,f) = \left(\frac{|S(t,f)|^2}{|S(t,f)|^2 + |N(t,f)|^2}\right)^\beta \quad (4)$$

where $\beta$ is a tunable parameter, which is usually set to 0.5.

#### C. Complex Ideal ratio mask (cIRM)

The IBM and the IRM are constructed on the magnitude of the speech and the noise. Recent research showed that the phase information is also important for speech separation [16]. The cIRM includes the phase in its construction, which can be viewed as the IRM in the complex domain.

Given the spectrum $Y$ of mixture, the spectrum $S$ of speech signal can be generated as follows,

$$S(t,f) = M(t,f) * Y(t,f) \quad (5)$$

where '*' is complex multiplication. $M(t,f)$ is the complex mask which can be expressed as follows,

$$M = \frac{Y_r S_r + Y_i S_i}{Y_r^2 + Y_i^2} + i \frac{Y_r S_i - Y_r}{Y_r^2 + Y_i^2} \quad (6)$$

where $Y_r$, $Y_i$, $S_r$ and $S_i$ stand for the real and imaginary components of $Y$ and $S$, respectively.

#### D. Phase sensitive mask (PSM)

The PSM directly uses a phase sensitive target function, which involves both amplitude and phase information.

$$\text{PSM}(t,f) = \frac{|S(t,f)|}{|Y(t,f)|} \cos\theta \quad (7)$$

where $\theta = \theta^S + \theta^Y$ is the phase. PSM is restricted in real domain.

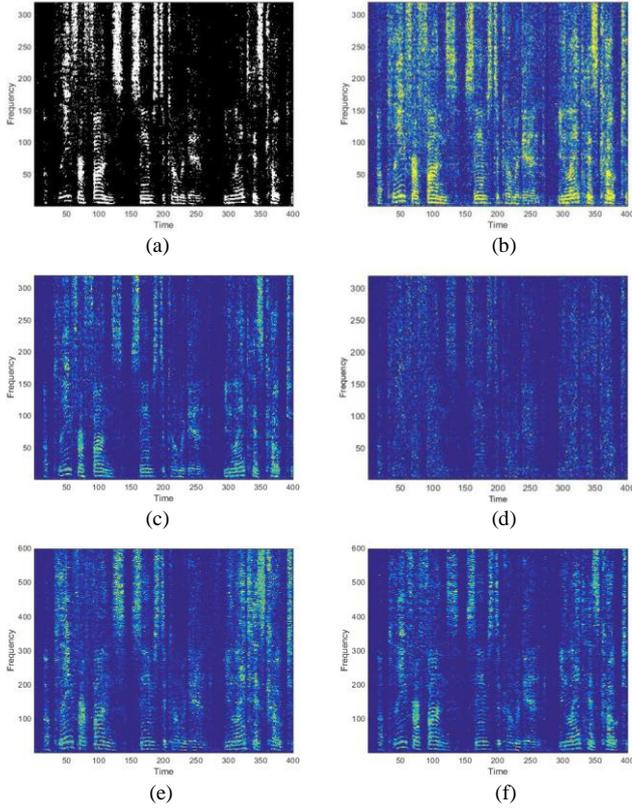

Fig. 3. Illustration of various mask targets. (a) is IBM; (b) is IRM; (c) and (d) are the real and imaginary parts of the cIRM; (e) is the PSM; (f) is the ORM.

## V. EXPERIMENTAL RESULTS

### A. Dateset

We use 600 randomly chosen utterances from the IEEE female corpus as our training utterances. The rest of 120 utterances are used as the test set. Four noises from the NOISEX dataset [17] are used as our training and test noises, including a speech-shaped noise (SSN), a babble noise (babble), a factory noise (factory) and a destroyer engine room noise (engine). Except SNN, all other 3 noises are non-stationary. All noises are around 4 minutes long. To create the training set, we use random 10 slices from the first 2 minutes of each noise to mix with each utterance from the training utterances at -3, 0 and 3dB. Thus, we have 72000 (600 utterances×10 slices×4 noises×3 SNR). The test mixtures are constructed by mixing random cuts from the last 2 minutes of each noise with the test utterances at -3, 0 and 3 dB. Cutting from different parts of the noises ensures that noise segments used in training and testing phase are different.

### B. Evaluation Criteria

We use the Short-Time Objective Intelligibility (STOI) score [18] to measure the objective intelligibility. STOI denotes a correlation of short-time temporal envelopes between clean and separated speech, and has been proved to be highly correlated to human speech intelligibility score. The value of STOI is in the range of [0, 1]. We also evaluate objective speech quality using the Perceptual Evaluation of Speech Quality (PESQ) score [19]. Same as the STOI, the PESQ is calculated by comparing the separated speech with the corresponding clean speech. The STOI score ranges from 0 to 1, and PESQ score ranges from -0.5 to 4.5.

### C. Results

The separating results of the five training targets is given in table 1, 2, 3, which are respectively under the condition of -3dB, 0dB, 3dB SNR mixture. Bold face indicates the target that performed best within a noise type. IBM and IRM are two most widely used training targets. Table 1~3 show that IBM improves speech intelligibility but not speech quality. This may due to the binary property of IBM, and musical noise is produced during the separation. Compared to IBM, IRM improves both speech intelligibility and speech quality significantly, especially the speech quality.

TABLE I
PERFORMANCE COMPARISONS BETWEEN VARIOUS TARGETS ON -3 dB MIXTURES

| Targets | PESQ | | | | STOI | | | |
|---|---|---|---|---|---|---|---|---|
| | SNN | Babble | Factory | Engine | SNN | Babble | Factory | Engine |
| Mixture | 1.37 | 1.53 | 1.56 | 1.39 | 0.61 | 0.60 | 0.65 | 0.60 |
| IBM | 1.19 | 0.84 | 1.65 | 1.09 | 0.69 | 0.63 | 0.81 | 0.68 |
| IRM | 2.05 | 1.80 | 2.30 | 1.94 | 0.77 | 0.70 | 0.86 | 0.75 |
| cIRM | 2.14 | 1.76 | 2.45 | 2.04 | 0.76 | 0.70 | 0.85 | 0.74 |
| PSM | 2.23 | **1.95** | 2.56 | 2.15 | **0.78** | **0.72** | **0.87** | **0.76** |
| ORM | **2.29** | 1.93 | **2.63** | **2.20** | 0.77 | 0.71 | 0.86 | 0.75 |

TABLE II
PERFORMANCE COMPARISONS BETWEEN VARIOUS TARGETS ON 0 dB MIXTURES

| Targets | PESQ | | | | STOI | | | |
|---|---|---|---|---|---|---|---|---|
| | SNN | Babble | Factory | Engine | SNN | Babble | Factory | Engine |
| Mixture | 1.53 | 1.73 | 1.67 | 1.57 | 0.68 | 0.67 | 0.71 | 0.67 |
| IBM | 1.52 | 1.23 | 1.87 | 1.43 | 0.78 | 0.74 | 0.86 | 0.77 |
| IRM | 2.32 | 2.11 | 2.61 | 2.09 | 0.83 | 0.79 | 0.90 | 0.82 |
| cIRM | 2.42 | 2.13 | 2.64 | 2.32 | 0.83 | 0.79 | 0.89 | 0.82 |
| PSM | 2.53 | 2.28 | 2.73 | 2.43 | **0.85** | **0.80** | **0.90** | **0.83** |
| ORM | **2.58** | **2.30** | **2.80** | **2.49** | 0.84 | **0.80** | **0.90** | **0.83** |

TABLE III
PERFORMANCE COMPARISONS BETWEEN VARIOUS TARGETS ON 3 dB MIXTURES

| Targets | PESQ | | | | STOI | | | |
|---|---|---|---|---|---|---|---|---|
| | SNN | Babble | Factory | Engine | SNN | Babble | Factory | Engine |
| Mixture | 1.72 | 1.91 | 1.80 | 1.73 | 0.74 | 0.74 | 0.78 | 0.74 |
| IBM | 1.76 | 1.64 | 2.08 | 1.73 | 0.84 | 0.83 | 0.90 | 0.84 |
| IRM | 2.56 | 2.40 | 2.77 | 2.39 | 0.88 | 0.86 | 0.92 | 0.87 |
| cIRM | 2.67 | 2.45 | 2.86 | 2.58 | 0.88 | 0.86 | 0.92 | 0.87 |
| PSM | 2.74 | 2.55 | 2.90 | 2.67 | **0.89** | **0.87** | **0.93** | **0.88** |
| ORM | **2.81** | **2.60** | **2.95** | **2.73** | **0.89** | **0.87** | 0.92 | **0.88** |

cIRM, PSM and ORM are training targets proposed recent years whose value various in a large range. cIRM performs best theoretically, the result shows it outperforms IRM on improvement of speech quality, whereas its improvement of speech intelligibility is close to IRM. cIRM and PSM both concerned about phase information, cIRM operates on complex domain while PSM on real domain. Table 1~3 show that PSM improves speech intelligibility by 1%~2% and 12%~22% compared to cIRM and unprocessed mixture, outperforms other training targets. PSM improves speech quality by 0.07~0.19 compared to cIRM. This maybe because that structure of the imagine component of cIRM is not obvious and difficult to estimate, which lead to its poor actual performance.

Table I, II and III show that the improvement on speech intelligibility of ORM and PSM are close, while ORM outperforms PSM on speech quality. We can see from Table I that improvement of ORM is 1% lower than PSM while better than other training targets. For the speech quality, on SSN, Engine and Factory noise, ORM performs best, which improves 0.81~1.07 compared to unprocessed mixture and 0.05~0.07 compared to PSM. In Table II, performance of ORM is close to PSM on speech intelligibility. ORM outperforms other training targets on speech quality. In Table III, on engine noise PSM is little bit better than ORM, while on other noise their performance are close as to speech intelligibility ORM outperforms other training targets. In general, ORM outperforms other training targets. ORM concerns about the correlation between clean speech and noise while PSM concerns about the phase information. ORM performs better than PSM may due to that the correlation between clean speech and noise have greater impact on speech separation than phase information, the estimated target signal is closer to clean speech. Fig 5 shows STFT spectrogram of mixture on Babble noise at 3dB and target speech separated with IBM, IRM, cIRM, ORM, PSM as the training targets.

## VI. Conclusions

Monaural speech separation is always a challenging task in the area of speech recognition. Supervised speech separation integrates deep neural network technology which emerged as a new trend in recent years, impressive improvements are achieved under low SNR mixture and non-stationary noise condition. The IBM and IRM are the most popular training targets. However, IBM improves speech intelligibility but not speech quality. IRM improves both speech intelligibility and speech quality under the assumption that clean speech is independent with noise. Recent research prove that phase information is important for speech separation, cIRM and PSM were proposed. cIRM performs well theoretically, while the structure of its imagine component is not obvious, difficult to estimate. PSM improves speech intelligibility and speech quality significantly, outperform other training targets.

In this paper, we adapt ORM as training target, which concerns about the correlation between clean speech and noise, the result shows that ORM performs better than other training targets in general. This maybe because that there is relation between clean speech and noise in real environment and have greater impact on speech separation than phase information. Based on this, we think the analyses of the relation between clean speech and noise and how to describe and estimate it is going to be a new direction in the area of training target research.

## Acknowledgment

This research was supported in part by a China National Nature Science Foundation grant (No. 61365006),